\documentclass[12pt]{iopart} % 10pt is ignored!
\usepackage{epsfig}                   % please use epsfig for figures

\begin{document} 
\title{Some comments on the missing charm puzzle}

\author{Alexander Lenz}
\address{Universit{\"a}t Regensburg, D-93040 Regensburg, Germany}
\ead{alexander.lenz@physik.uni-regensburg.de}

\begin{abstract}
       In this talk we summarize the status of theoretical 
       predictions for the average number of charm quarks in a B-hadron
       decay.
\end{abstract}
%\maketitle
%-------------------------------------------------------------------------
\section{Introduction}
%-------------------------------------------------------------------------
%-------------------------------------------------------------------------
Since quite a long time there exists a discrepancy between theoretical 
predictions and measurements of the quantity $n_c$, which describes the 
average number of charm quarks in the final state of a B-hadron decay 
\cite{nchist}.
In the last years this difference became smaller and it became a matter 
of taste whether one speaks of a missing charm puzzle or not. In this talk 
we try to summarize the theoretical results and to clarify the origin of 
different numbers for $n_c$.
\\
One can calculate $n_c$ in the following ways:
\begin{eqnarray}
n_c 
& = & 
0 + 
\frac{\Gamma ( b \to 1 c)}{\Gamma_{tot}}
+ 2 
\frac{\Gamma ( b \to 2c)}{\Gamma_{tot}}
\\
&& 
\nonumber
\\
& = & 
1 
+ 
\frac{\Gamma ( b \to 2c)}{\Gamma_{tot}}
-
\frac{\Gamma ( b \to 0c)}{\Gamma_{tot}}
\\
&& 
\nonumber
\\
& = & 
2 
- 
\frac{\Gamma ( b \to 1c)}{\Gamma_{tot}}
-
2
\frac{\Gamma ( b \to 0c)}{\Gamma_{tot}}
\end{eqnarray}
$\Gamma (b \to 0c)$ sums up all charmless decay rates like
the non-leptonic channels
$ b \to u \bar{u} s,d$, $ b \to s \bar{s} s,d$, $ b \to d \bar{d} s,d$ and
the semi-leptonic channels $ b \to u l \nu$ and $ b \to s g,gg$.
$\Gamma (b \to 1c)$ sums up all decay rates with one charm 
quark in the final state, like
the non-leptonic channels
$ b \to c \bar{u} s,d$, $ b \to u \bar{c} s,d$ and 
the semi-leptonic channels $ b \to c l \nu$. 
Finally we have $\Gamma (b \to 2c)$ with two charm quarks 
in the final state: $ b \to c \bar{c} s,d$.
\\
Before we compare experimental results and theoretical predictions, let us
look at the calculation of these decay rates.
%-------------------------------------------------------------------------
%-------------------------------------------------------------------------
\section{Calculation of inclusive decay rates}
%-------------------------------------------------------------------------
%-------------------------------------------------------------------------
The Heavy Quark Expansion (HQE) (for a recent review see \cite{hqe}) 
is the theoretical framework to handle
inclusive $B$-decays. It allows us to
expand the decay rate in the following way
\begin{equation}
\Gamma  =  \Gamma_0 
+ \left( \frac{\Lambda}{m_b} \right)^2 \Gamma_2
+ \left( \frac{\Lambda}{m_b} \right)^3 \Gamma_3
+ \cdots 
\label{hqe}
\end{equation}
Here we have an systematic expansion in the small parameter
$\Lambda/m_b$. The different terms have the following physical 
interpretations:
\begin{itemize}
\item $\Gamma_0$: The leading term is described by the decay of a free
                  quark (parton model),
                  we have no non-perturbative corrections.

\item $\Gamma_1$: In the derivation of eq. (\ref{hqe}) we make an 
                  operator product expansion. From dimensional reasons
                  we do not get an operator which would contribute to
                  this order in the HQE. \footnote{Strictly spoken we
                  get one operator of the appropriate dimension, 
                  but with the equations of motion we can
                  incorporate it in the leading term.}
      
\item $\Gamma_2$: First non-perturbative corrections arise at the
                  second order in the expansion due to the
                  kinetic and the chromomagnetic operator. They can be
                  regarded as the first terms in a non-relativistic
                  expansion.

\item $\Gamma_3$: In the third order we get the so-called weak
                  annihilation and pauli interference diagrams. 
                  Here the spectator quark is included for the first
                  time.
                  These diagrams give rise to different lifetimes 
                  for different $B$ hadrons.
\item             The dots represent higher order terms in $1/m_b$, possible 
                  non-perturbative $1/m_c^2$ corrections (like in the decay
                  $ B \to X_s \gamma$ \cite{1/mc}) and
                  unknown terms which are due to duality violation 
                  (see  \cite{shifman} for a nice review).
\end{itemize}
Schematically one can write the $\Gamma_i$'s as products of perturbatively
calculable functions 
(depending on couplings, masses, renormalization scale,...) and matrix 
elements, which have to be determined by some non-perturbative methods 
like lattice-QCD or sum rules. 
\\
Now we may have a closer look at eq. (\ref{hqe}).
Each of the appearing terms can be expanded in a power series in the
strong coupling constant
\begin{equation}
\Gamma_i =  \Gamma_i^{(0)} + 
\frac{\alpha_s}{\pi}   \Gamma_i^{(1)} + \cdots \; .
\end{equation}
We start with a discussion of the perturbative part of the 
$\Gamma_i^{(j)}$'s and then we make some comments about the status of
the non-perturbative parameters.
%-------------------------------------------------------------------------
\subsection{Leading term: $\Gamma_0$}
%-------------------------------------------------------------------------
%\begin{displaymath}
%\Gamma_0 =  \Gamma_0^{(0)} + 
%\frac{\alpha_s}{\pi}   \Gamma_0^{(1)} + \cdots \; .
%\end{displaymath}
%What has been done?
 $\Gamma_0^{(0)}$ is well known. 
 In addition we have analytic expressions
 of $\Gamma_0^{(1)}$ for $ b \to c l \nu$ \cite{nir}
 and $ b \to c \bar{u} d$ \cite{ball1} and a 
 numerical value for $ b \to c \bar{c} s$ \cite{ball2}. 
 The effects of the charm quark mass were found to be quite sizeable.
 Although suppressed by one power of $\alpha_s$, penguin diagrams are 
 dominant for $ b \to no \; charm$ \cite{lenz1}, \cite{lenz2}. 
 Recently the NLO calculation for
 $ b \to s g $ has been finished \cite{greub1}.
%What has to be done
 The inclusion of penguin diagrams with current-current operators for 
 the decay $ b \to c \bar{c} s$
 and penguin diagrams with penguin operators for $ b \to no \; charm$
 is still missing, but their effects are not expected to be large.
%Matrix elements
 It is a remarkable feature of the HQE that in the leading term  $\Gamma_0$
 only the unit operator appears, so the matrix elements of this operator 
 are trivial. Therefore we have no non-perturbative parameters in 
 $\Gamma_0$.
%-------------------------------------------------------------------------
\subsection{Sub-leading term: $\Gamma_2$}
%-------------------------------------------------------------------------
%\begin{displaymath}
%\Gamma_2 =  \Gamma_2^{(0)} + 
%\frac{\alpha_s}{\pi}   \Gamma_2^{(1)} + \cdots \; .
%\end{displaymath}
%What has been done?
 $\Gamma_2^{(0)}$ is known for the most important operator insertions
 \cite{1/m_b^2}. 
%What has to be done
 Some penguin operator insertions are still missing. It would be nice to
 have a result for $\Gamma_2^{(1)}$, but the calculation seems to be 
 quite tough.
 One has to calculate the imaginary part of three loop diagrams with one
 external gluon.
%Matrix elements
 Here we have two matrix elements: $\lambda_1$ and $ \lambda_2$.
 The first one is not very well known, see e.g \cite{talkbraun}, while the 
 second number can be extracted from experiment. 
%-------------------------------------------------------------------------
\subsection{Spectator effects: $\Gamma_3$}
%-------------------------------------------------------------------------
Spectator effects arise first in the third order of the expansion in $1/m_b$.
%\begin{displaymath}
%\Gamma_3 =  \Gamma_3^{(0)} + 
%\frac{\alpha_s}{\pi}   \Gamma_3^{(1)} + \cdots \; .
%\end{displaymath}
%What has been done?
 $\Gamma_3^{(0)}$ is known for $\Delta \Gamma_{B_S} $\cite{dgbslo}
 and for $B^+$, $B_s$ and $\Lambda_b$ with charm quark mass effects
 \cite{neubert1}.
 $\Gamma_3^{(1)}$ was calculated for $\Delta \Gamma_{B_S} $ by 
 \cite{dgbsnlo}.
%What has to be done ?
 The calculation of $\Gamma_3^{(1)}$ for $B^+$, $B_s$ and $\Lambda_b$
 is still missing.
%Matrix elements
 In $\Gamma_3$ we have the following non-perturbative parameters: decay 
 constants $f_M$ (depending on the decaying meson $M$)
 and Bag-Barameters $B_{D_M}$ (depeding on the decaying meson $M$ and the 
 Dirac structures $D$ of the appearing operators). 
 For $\Delta \Gamma_{B_S} $ we have already quite stable lattice predictions
 for these quantities, while for
 $B^+$, $B_s$ and $\Lambda_b$ relieable numbers are still missing 
 (see \cite{durhambeneke}, \cite{durhamflynn}).
%-------------------------------------------------------------------------
\subsection{$1/m_b^4$ corrections: $\Gamma_4$}
%-------------------------------------------------------------------------
%\begin{displaymath}
%\Gamma_4 =  \Gamma_4^{(0)} + 
%\frac{\alpha_s}{\pi}   \Gamma_4^{(1)} + \cdots \; .
%\end{displaymath}
%What has been done?
 For $\Delta \Gamma_{B_S} $ even 
 $\Gamma_4^{(0)}$ has been calculated by \cite{bbd};
%What has to be done ?
 This could be done for $B^+$, $B_s$ and $\Lambda_b$, too.
%Matrix elements
 The appearing matrix elements were estimated in vacuum insertion 
 approximation.
%-------------------------------------------------------------------------
%-------------------------------------------------------------------------
\section{Different normalization}
%-------------------------------------------------------------------------
%-------------------------------------------------------------------------
In order to determine $n_c$ we have to determine the branching ratios
for $b$ decays into 0,1 and 2 charm quarks. 
So one could simply calculate $ \Gamma (0,1,2c)$ and $ \Gamma_{tot}$.
But there are several reasons, why it 
might be better not to calculate these quantities straightforward.
\\
First, the semi-leptonic decay rate $\Gamma_{sl}$ is clearly the most reliable 
prediction, while $\Gamma_{tot}$ is probably the least reliable prediction.
By writing
\begin{equation}
B_{b \to X} = \frac{\Gamma_X}{\Gamma_{sl}}
* 
\frac{\Gamma_{sl}}{\Gamma_{tot}} 
=: r_X * B_{sl}^{exp}
\label{rratio}
\end{equation}
we can eliminate $\Gamma_{tot}$ in favor of $\Gamma_{sl}$. 
In $r_X$ we have no $m_b^5$- and $\lambda_{1}$-dependence anymore.
\\
Second, the decay $ b \to c \bar{c} x$ is most sensitive to possible
quark hadron duality violations. This is due to the fact that the HQE is 
actually not an expansion in $1/m_b$, but in $1/E$, where E is the energy 
release in the decay. For $ b \to c \bar{c} x$ we have $ E = m_b - 2 m_c$,
which is already quite a small number. If we use eq. (3) and the $r$'s
instead of the branching ratios, we have eliminated the decay 
$ b \to c \bar{c} x$, as proposed in \cite{bdy}.
Now $ r(0c) $ is an important input parameter for the determination of $n_c$.
Possible enhancements of $ r(0c) $ due to new physics would lower
$ B_{sl}^{theory}$ and $ n_{c}^{theory}$ simultaneously. Different mechanisms
for such an enhancement were studied in the literature \cite{newphysics}.
%-------------------------------------------------------------------------
%-------------------------------------------------------------------------
\section{Results in the literature}
%-------------------------------------------------------------------------
Now we summarize the results for the relevant decay rates from the literature 
and determine $n_c$ in various ways. 
%-------------------------------------------------------------------------
%-------------------------------------------------------------------------
\subsection{Counting of one Charm Quark}
%-------------------------------------------------------------------------
      The dominant decay is $ b \to c \bar{u}d$. There was quite a 
      confusion due to two different numbers in the literature:
      Ball et al. quote $r (c \bar{u}d) =  4.0 \pm 0.4$\cite{ball1},
      while Neubert was showing 
      $r (c \bar{u}d) =  4.2 \pm 0.4$\cite{neubertjeru} in Jerusalem.
      The difference of these numbers is an effect of second order in 
      $\alpha_s$. While the authors of \cite{ball1} were calculating ratios 
      like $(a + \alpha_s b)/(c + \alpha_s d)$ nummerically, the author of 
      \cite{neubertjeru} expanded the ratio in $\alpha_s$ \cite{private}. 
      Unfortunateley the difference is quite sizeable.
      For all possible semi-leptonic decays we get 
      $r_{c l \nu}  =  2.22 \pm 0.04$ 
      and for the Cabibbo suppressed decay modes the result is
      $ r_{u \bar{c} s'}  =  0.03 \pm 0.00$.
Depending on our input for $r (c \bar{u}d)$ we get two different results:
\begin{displaymath}
r (1c) = 6.25 \pm 0.4 \; \cite{ball1}
\hspace{3cm}
r (1c) = 6.45 \pm 0.4 \; \cite{neubertjeru}
\end{displaymath}
%-------------------------------------------------------------------------
\subsection{Counting of no Charm Quark}
%-------------------------------------------------------------------------
For the non-leptonic charmless $b$-decays it turned out, that penguin 
diagrams are as important as the leading contribution to these decays, 
although being suppressed by $\alpha_s$ \cite{lenz1}.
Even $\alpha_s^2$ contributions, so-called double penguins have a sizeable
value \cite{lenz2}. One gets $r (0c) = 0.18 \pm 0.08$ \cite{lenz1,lenz2} 
for all charmless final states.
Recently the NLO QCD calculation of $ b \to sg $ and $ b \to sgg$ was 
finished \cite{greub1}. Greub and Liniger get an enhancement of more than
$100 \%$ compared to the LO value
\begin{displaymath}
r (b \to sg,sgg) = \left\{
\begin{array}{cc}
0.022 \pm 0.008 & \mbox{LO}   
\\
0.05 \pm 0.01   & \mbox{NLO} \; .
\\
\end{array}
\right. 
\end{displaymath}
With the new result for $ b \to sg $ and $ b \to sgg$ at hand we get:
\begin{displaymath}
r (0c) = 0.21 \pm 0.08 \; \cite{greub1}
\end{displaymath}
%-------------------------------------------------------------------------
\subsection{Counting of two Charm Quarks}
%-------------------------------------------------------------------------
For $ b \to c \bar{c} s$ we have again two different results. 
Ball et. al quote $r (2c)=2.0 \mp 0.5$ \cite{ball2}, while Neubert
gets $1.89 \mp 0.54$ \cite{neubertjeru}.
The difference has the same origin as in section 4.1. 
%-------------------------------------------------------------------------
\subsection{Results for $n_c$}
%-------------------------------------------------------------------------
With the experimental value for the semi-leptonic branching ratio
presented in Osaka $B_{sl}^{exp.} = 0.1059 \pm 0.0016$ \cite{barker},
we can determine $n_c$ in three different ways.
\begin{enumerate}
\item Elimination of no charm: $n_c 
=  \left(r(1c) + 2 r(2c) \right) B_{sl}^{exp} 
= 1.09 \pm 0.11$
\item Elimination of one charm: $n_c 
= 1 + \left(r(2c) - r(0c) \right) B_{sl}^{exp} 
= 1.18 \pm 0.06$
\item Elimination of two charm: $n_c 
= 2- \left(r(1c) + 2 r(0c) \right) B_{sl}^{exp} 
= 1.28 \pm 0.05$
\end{enumerate}
For $r(1c)$ and $r(2c)$ we used the average of \cite{ball1, ball2} and
\cite{neubertjeru}.
Of course, all these numbers should be the same. The reason for the
disagreement is found by comparing the theoretical and experimental value
of the semi-leptonic branching ratio.
Theory tells us
\begin{displaymath}
\left( r(0c) + r(1c) + r(2c) \right)^{-1}
= 0.118 \pm 0.009 = B_{sl}^{theory} \; .
\end{displaymath}
The central value is quite above the experimental number for $B_{sl}$, but
the errors are large. 
When we introduced $r_X$ in eq. (\ref{rratio}), we asummed 
that $B_{sl}^{theory}=B_{sl}^{exp.}$, which is not satisfied. This is the 
reason for the inconsistencies in the determination of $n_c$.
If we use $B_{sl}^{theory}$ to determine $n_c$, we get in all three cases
the central value $ n_c = 1.21$.
\\
In Osaka $ n_c = 1.16 \pm 0.05$ was given as the experimental value
\cite{barker}, while
Kagan gets a value of $ n_c = 1.085 \pm 0.05$ \cite{kaganprivate}.
It is beyond the scope of this talk to clarify the origin of these two 
different experimental numbers .
%-------------------------------------------------------------------------
\section{Disscussion and outlook}
%-------------------------------------------------------------------------
%-------------------------------------------------------------------------
In this talk we tried to clarify the orgin of different values for $n_c$
on the market.
First we have different numbers for $r(1c)$ and $r(2c)$ due to a different
treatment of ${\cal O}(\alpha_s^2)$ contributions. The numbers of 
\cite{neubertjeru} give a slightly smaller value for $n_c$, than the numbers
of \cite{ball1, ball2}.
Second, we get quite different results for the three possibilities 
(eq. (1)-(3)) to determine $n_c$, if we use a normalization of the decay 
rates to $\Gamma_{sl}$ instead of $\Gamma_{tot}$\footnote{In the 
determination of $\Delta \Gamma_{B_s}$ we have the 
same situation, that we get quite different numbers for different 
normalizations (see talk \cite{durhambeneke}).}.
The reason for that is the disagreement of the theoretical number for 
$B_{sl}$ with the experimental value. This problem has to be resolved 
in the future.
Third, the experimental value of $n_c$ seems to be not completely clear. 

So we are still not in the position to say the final word about the existence 
of a missing charm puzzle. If we use an appropriate theoretical 
input and set $ \mu = m_b/4$ (which means a high value for $\alpha_s$)
and $ m_c/m_b = 0.33$\footnote{Here one should keep in mind, that the 
ratio $ m_c/m_b $ is fixed by HQET.}, than experiment 
(the numbers shown in Osaka) and theory agree more or less. 
On the other hand there is still room for a deviation, which might be due to
a new physics enhanced $r(0c)$ or dualtity violation in 
{$ b \to c \bar{c} s$} or....
Precise experimental values of $ r(2c) $ and $ r(0c) $ would help a lot, 
to confirm or to rule out these interesting possibilities.
%-------------------------------------------------------------------------

\ack
I want to thank the organizers of the workshop
for their successful work, U. Nierste for the pleasant
collaboration, V. Braun, P. Ball, C. Greub, A. Kagan, P. Liniger, 
M. Neubert and N. Uraltsev for useful informations and discussions 
and V. Braun and M. Schenk for proofreading the manuscript.

\end{document}